\documentclass[aps,prd,twocolumn,showpacs,groupedaddress]{revtex4}

\usepackage{graphicx}
\usepackage{color}

\def\apj #1 #2 #3 {#1, ApJ, {\bf #2}, #3}
\def\apjl #1 #2 #3 {#1, ApJ, {\bf #2}, L#3}
\def\apjs #1 #2 #3 {#1, ApJS, {\bf #2}, #3}
\def\aap  #1 #2 #3 {#1, A\&A, {\bf #2}, #3}
\def\mnras #1 #2 #3 {#1, MNRAS, {\bf #2}, #3}
\def\pra #1 #2 #3 {#1, Phys.~Rev.~A., {\bf #2}, #3}
\def\prb #1 #2 #3 {#1, Phys.~Rev.~B., {\bf #2}, #3}
\def\prc #1 #2 #3 {#1, Phys.~Rev.~C., {\bf #2}, #3}
\def\prd #1 #2 #3 {#1, Phys.~Rev.~D., {\bf #2}, #3}
\def\pre #1 #2 #3 {#1, Phys.~Rev.~E., {\bf #2}, #3}
\def\prl #1 #2 #3 {#1, Phys.~Rev.~Lett., {\bf #2}, #3}
\def\plb #1 #2 #3 {#1, Phys.~Lett.~B., {\bf #2}, #3}
\def\science #1 #2 #3 {#1, Science., {\bf #2}, #3}
\def\nature #1 #2 #3 {#1, Nature., {\bf #2}, #3}
\def\nphysa #1 #2 #3 {#1, Nucl.~Phys.~A., {\bf #2}, #3}
\def\nphysb #1 #2 #3 {#1, Nucl.~Phys.~B., {\bf #2}, #3}
\def\nphysbs #1 #2 #3 {#1, Nucl.~Phys.~B.~Suppl., {\bf #2}, #3}

\def\h#1{\hbox{${}^{#1}$H}}

\def\h502{\hbox{$ h^{2}_{50}$}}

\def\fun#1#2{\lower3.6pt\vbox{\baselineskip0pt\lineskip.9pt
  \ialign{$\mathsurround=0pt#1\hfil##\hfil$\crcr#2\crcr\sim\crcr}}}
\begin{document}
\bigskip
\bigskip
%

\title{Exploring the Neutrino Mass  Hierarchy Probability with Meteoritic  Supernova Material, $\nu$-Process Nucleosynthesis,  and  $\theta_{1 3}$ Mixing}
\author{
G. J. Mathews,$^{1}$
T. Kajino,$^{2,3}$
W. Aoki,$^2$
W. Fujiya,$^4$
J. B. Pitts$^5$ 
}
\address{$^1$University of Notre Dame, Center for Astrophysics, Notre Dame,
IN 46556
}  
\address{$^2$National Astronomical Observatory, Mitaka, Tokyo 181-8588, Japan
}
\address{
$^3$Department of Astronomy, Graduate School of Science,
University of Tokyo, 7-3-1
Hongo, Bunkyo-ku, Tokyo 113-0033, Japan }
\address{$^4$Department of Earth and Planetary  Science, 
University of Tokyo, 7-3-1 Hongo, Bunkyo-ku, Tokyo, 113-0033, Japan
}
\address{$^5$Departments of Physics and Philosophy, University of Notre Dame, Notre Dame, IN 46556
}


\date{\today}
\begin{abstract}
There is  recent evidence that some SiC X grains from the Murchison  meteorite may contain supernova-produced $\nu$-process  $^{11}$B  and or $^7$Li encapsulated in the grains. The synthesis  of $^{11}$B and $^{7}$Li via neutrino-induced nucleon emission (the $\nu$-process) in supernovae  is sensitive to the neutrino mass hierarchy for finite  $\sin^2{2 \theta_{1 3}} > 0.001$.   This  sensitivity arises because, when there is 13 mixing, the average electron neutrino energy for charged-current neutrino reactions is  larger for a normal mass hierarchy than for an inverted hierarchy. 
Recent constraints on  $\theta_{1 3}$ from the Daya Bay, Double Chooz, MINOS, RENO and T2K  collaborations all suggest that  indeed $\sin^2{2 \theta_{1 3}} > 0.001$.    We examine the possible implications of these new results based upon a Bayesian analysis of  the uncertainties in the measured meteoritic material and the  associated supernova nucleosynthesis models.  We show that although the uncertainties are large,  they  hint at a marginal  preference for an inverted neutrino mass hierarchy.  We discuss the possibility  that an analysis of  more X grains enriched in Li and B  along with a  better understanding of the relevant stellar nuclear and neutrino reactions could eventually reveal the neutrino mass hierarchy.
\end{abstract}
%
%



\maketitle

\section{Introduction}
Oscillations in the three-neutrino flavor mixing scenario are described by three angles $\theta_{1 2}$, $\theta_{2 3}$, and $\theta_{1 3}$ plus a CP-violating phase $\delta_{CP}$.  At the present time solar, atmospheric, and reactor  neutrino oscillation  measurements  \cite{SuperK05,SuperK06,SNO,K2K,KamLAND,MINOS} have  provided information on the neutrino mass differences, i.e $\Delta m_{1 2}^2 \equiv |m_1^2 - m_2^2| = 0.000079$ eV \cite{KamLAND} and $\Delta m_{1 3}^2 \approx \vert  \Delta m_{23}^2 \vert \approx 2.4 \times 10^{-3}$ eV$^2$ \cite{MINOS, SuperK06}.  These measurements, however,  are unable to determine the mass hierarchy, i.e. whether $\Delta m^2_{23} > 0$ (normal) or $\Delta m^2_{23} <  0$ (inverted) is the correct order.  Also,  solar, atmospheric and reactor neutrino oscillation experiments \cite{SuperK05,SuperK06,SNO,K2K,KamLAND,MINOS} have determined $\theta_{1 2}$ and $\theta_{2 3}$ to reasonable precision.  However, only recently have measurements of   $\theta_{1 3}$ become available.  The three best best current measurements [with $\delta_{CP}  = 0$ and $\theta_{2 3} = \pi/4$] are  from the the Daya Bay experiment that  has recently  reported \cite{DayaBay}  $\sin^2{2 \theta_{1 3}}= 0.092 \pm 0.016 ({\rm stat}) \pm 0.005 (\rm syst)$,  
 the  independent recent report \cite{RENO} from the RENO collaboration of $\sin^2 2 \theta_{13} = 0.113 \pm 0.013({\rm stat.}) \pm 0.019({\rm syst.})$ and the value \cite{Chooz} from the Double Chooz collaboration  of $\sin^2 2 \theta_{13} = 0.086 \pm 0.041({\rm stat.}) \pm 0.030({\rm syst.})$.  
 
 These results are mutually consistent and also consistent with the previously reported upper limit from the MINOS collaboration \cite{MINOS} of $ \sin^2{2 \theta_{1 3}} < 0.12 (0.20)$  for  the normal (inverted) hierarchy,  and  with  the  90\% C.L.~lower limit and upper limits  to $\theta_{1 3}$ from the T2K collaboration \cite{T2K} of $0.03 (0.04) < \sin^2{2 \theta_{1 3}} < 0.28 (0.34)$.  As encouraging as these new results are, however, the data do not yet determine whether the normal or inverted hierarchy is the correct ordering of mass eigenstates.

In this context we note that previous studies \cite{Yoshida04, Yoshida05, Yoshida06a, Yoshida06b,Yoshida08} of the nucleosynthesis of the light isotopes via the $\nu$-process in  supernovae have pointed out  that for  a finite mixing angle $\theta_{1 3} > 0.001$ the relative synthesis of $^7$Li and $^{11}$B in the $\nu$-process is sensitive to the mass hierarchy.  Since $\theta_{1 3} > 0.001$ is indeed implied  by  the Daya Bay +RENO + Double Chooz results to better than the $5 \sigma$ C.L., it is worthwhile to reconsider the supernova $\nu$-process as a  means to constrain the neutrino mass hierarchy.  Moreover,  there has also recently appeared a possible discovery \cite{Fujiya11} of $\nu$-process supernova material in SiC X grains from the Murchison meteorite.  The goal of this paper is to examine whether constraints on the neutrino mass hierarchy can be determined from these data.

  It has recently been pointed out \cite{Austin11}, however, that the supernova models themselves can lead to variations in the  synthesis of $^7$Li and $^{11}$B that are much larger than the anticipated mass hierarchy effect due to uncertainties in the stellar thermonuclear reaction rates.   Moreover, the previous studies only considered a single 16.2 M$_\odot$  supernova model, while there is a significant dependence of the $\nu$-process yields on supernova progenitor mass which can vary  \cite{Heger05,Woosley95} from 10 to 25 M$_\odot$.  Nevertheless, rather than to give up on this quest, one should keep in mind that there are prior constraints on the degree to which the nuclear reaction rates can be varied, and on the probability for any particular supernova progenitor model.  Our goal here is to consider the likelihood of one neutrino mass hierarchy over another in a statistical analysis that takes proper account of all of these effects.  
  
  Indeed, the formulation of Bayesian statistics  \cite{HowsonUrbach,Gregory} provides just such a framework in which to analyze the probability that any particular model is true even in the context of uncertain data and large variations in underlying predictions due to model parameter uncertainties.  Moreover, this approach provides a means in which to incorporate all prior constraints on model parameters and input data.  In the remainder of this paper, therefore, we review the underlying model predictions, and their uncertainties along with an analysis of the meteoritic constraints.  These are then applied in a five dimensional Bayesian analysis.  We find that in spite of the large uncertainties, there remains a marginal preference for an inverted hierarchy at the level of 74\%/26\% compared to a prior expectation of a 50\%/50\%. 

\section {The $\nu$-process}
The $\nu$-process \cite{Domogatsky78, Woosley90} is believed to occur in core-collapse supernovae.  As the core collapses to high temperature and density,  neutrinos of all flavors are generated thermally in the proto-neutron star formed by the collapse.  These neutrinos emerge from the proto-neutron star  at relatively high temperature such that some neutrinos can participate in  neutrino-induced nucleon emission reactions.  The effects of these neutrino interactions are usually negligible.  However, the $\nu$-process can be a major contributor to the production of very rare isotopes not produced by other means such as
the  light isotopes,  $^7$Li, $^{11}$B,  $^{19}$F, and the heavy rare odd-odd  isotopes,  $^{138}$La and $^{180}$Ta \cite{Woosley90, Dighe00, Heger05, Hayakawa10}.  

The isotopes  $^7$Li and $^{11}$B, are of  particular interest for the present work.  These nuclides  can be copiously produced (e.g. $(^7$Li/H)/$(^7$Li/H)$_\odot \sim 10^4$ \cite{Yoshida06a,Yoshida06b,Yoshida08}) in supernovae as the neutrinos flow through the outer He layer of the exploding star.  Although $^7$Li can be also be produced in the Si shell \cite{Woosley90}, it will be   quickly destroyed by photodisintegration.  Therefore, in the present work we expect the final ejected abundances to be predominantly from the  He shell and can disregard production in the Si shell.  
Key to the present context is the sensitivity \cite{Yoshida08} of the produced $^{7}$Li/$^{11}$B abundance ratio to the neutrino mass hierarchy.  This arises because  neutrino oscillations can increase  the temperature characterizing the spectrum of reacting  neutrinos as
they  transport through the SN ejecta \cite{Dighe00}. This 
can increase the  rates
of  $\nu$-reactions  relative to  models without oscillations thus affecting the yields of the light elements.  

The dependence on $\theta_{13}$ enters because there is  a resonance of the 13-mixing in the
outer C/O layer \cite{Yoshida04, Yoshida05, Yoshida06a, Yoshida06b,Yoshida08}.  The associated increase  in the electron neutrino temperature increases the rates of charged-current $\nu$-process reactions in the outer He-rich layer.
The heavy  $\nu$-process elements are
 produced in the O-rich layers below the resonance region  \cite{Heger05} and  are thus not
 affected by the $13$ neutrino oscillations.   The light isotopes $^7$Li and $^{11}$B, however,
are significantly produced in the outer He layer as described below, and thus are sensitive to the 13 mixing.
Moreover, the produced  $^7$Li /$^{11}$B ratio depends not only on the mixing parameter, $\theta_{13}$,
but also on  the neutrino mass
hierarchy if the mixing angle is as large as suggested by current measurements.
  
The detailed dependence on the neutrino 
 mass hierarchy is as follows \cite{Yoshida04, Yoshida05, Yoshida06a, Yoshida06b,Yoshida08}. 
For a normal hierarchy, the  adiabatic  13-mixing
resonance for neutrinos in the O/C layer causes the $\nu_e$ energy spectrum in
the He/C layer to be  comparable to that of the $\nu_{\mu \tau}$ in 
the O-rich layer.  Without oscillations, $^7$Be is produced through the 
$^4$He$(\nu, \nu' n)^3$He$(\alpha,\gamma)^7$Be reaction sequence. However, when there are neutrino oscillations,
the rate of  the $^4$He$(\nu_e, e^- p)^3$He reaction exceeds the $^4$He$(\nu, \nu' n)^3$He  rate so that the production  of
$^7$Be in the He layer is much larger if neutrino oscillations occur.

 The production of $^7$Li is also greater when
 neutrino oscillations occur.  In this case, however,  the enhancement
is much less than that for $^7$Be. The main production
path for  $^7$Li is the $^4$He$(\nu, \nu' p)^3$H$(\alpha,\gamma)^7$Li reaction sequence and the corresponding
charged-current reaction is the $^4$He$(\bar \nu_e, e^+ n)^3$H reaction. 
There is, however,  no resonance for the antineutrinos, so this rate is insensitive to the 13 mixing.
 
 Neutrino oscillations also affect the production of  $^{11}$B and $^{11}$C similarly  to that of $^7$Li and
$^7$Be.  During
the $\nu$-process, $^{11}$C is normally produced via the  $^{12}$C$(\nu, \nu' n)^{11}$C reaction, plus a smaller contribution
from the  $^{12}$C$(\nu_e, e^- p)^{11}$C reaction.  When oscillations occur,
 the rate of the  $^{12}$C$(\nu_e, e^- p)^{11}$C
 becomes larger than
the rate  without oscillations by  about  an  order of magnitude so that the production  of $^{11}$C in the He layer is larger when neutrino oscillations
occur.

The abundance of $^{11}$B with oscillations, however,  is only slightly
larger when oscillations occur. The main production
process for $^{11}$B  is the $^4$He$(\nu, \nu' p)^3$H$(\alpha,\gamma)^7$Li$(\alpha,\gamma)^{11}$B reaction sequence.
The corresponding charged-current reaction is the $^4$He$(\bar \nu_e, e^+ n)^3$H reaction.
About 12\%-16\% of the $^{11}$B in the He layer
is also produced from $^{12}$C through  the $^{12}$C$(\nu, \nu' p)^{11}$B and the 
$^{12}$C$(\bar \nu_e, e^+ p)^{11}$B reactions. However, oscillations produce  no increase in the $^{11}$B production through
the $^{12}$C$(\bar \nu_e, e^+ n)^{11}$B or $^4$He$(\bar \nu_e, e^+ n)^3$H reactions,  because  the  antineutrinos
 have no resonance.   
 
 In the case of an inverted hierarchy, the production of $^7$Li
and $^{11}$B is greater than  for a normal hierarchy. The
$^4$He$(\bar \nu_e, e^+ n)^3$H and $^{12}$C$(\bar \nu_e, e^+ n)^{11}$B reaction 
 rates are  
larger because of an adiabatic resonance of $\bar \nu_e \leftrightarrow \bar \nu_{\mu \tau}$.
However, the increased production of  $^7$Li and
$^{11}$B is less pronounced than in the case of a 
normal hierarchy. This is because the average $\bar \nu_e$ energy is already greater 
 than the $\nu_e$ energy  at the neutrino sphere.  Therefore,  
 the average energy is not changed much by the oscillation with $\nu_{\mu \tau} (\bar \nu_{\mu \tau}$).
The mass fractions of $^7$Be and $^{11}$C are
slightly larger in the  outer parts of  the star, and slightly
smaller inside the He layer. There is, however,  no 13-mixing resonance for $\nu_e$.  Hence, there is no significant conversion of
 $\nu_e \leftrightarrow \nu_{\mu \tau}$.  At the same time, some of the produced  $^7$Be and
$^{11}$C is affected via capture neutrons from the $^4$He$(\bar \nu_e, e^+ n)^3$H reaction.

The final result from all of the above is that the emergent $^7{\rm Li_\nu}/{^{11}{\rm B_\nu}}$ ratio from the He layer is sensitive to whether the neutrino mass hierarchy is normal or inverted \cite{Yoshida06a, Yoshida06b, Yoshida08}. (We hereafter denote isotopes produced in the $\nu$-process by $^A$Z$_\nu$.) 

\begin{figure}[htb]
\includegraphics[width=3.8in,clip]{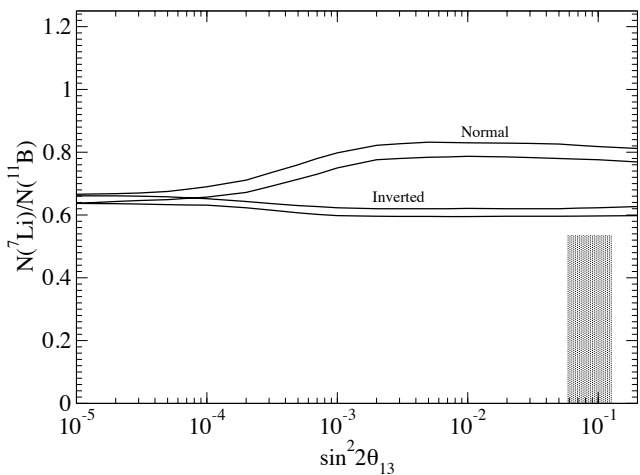} 
\caption{ Produced $^7$Li/$^{11}$B abundance \cite{Yoshida08} as a function of mixing angle for both a normal and inverted neutrino mass hierarchy.  This ratio varies  in models with different neutrino temperatures in the range indicated by the lower  and  upper solid lines.  The  width of the shaded region indicates the $2\sigma$ confidence  limits to  $\sin^2{(2\theta_{1 3})}$ from the  Daya Bay result \cite{DayaBay}.  The top of the shaded region is the $2 \sigma$ upper limit on the  observed $\nu$-process $^7$Li/$^{11}$B ratio as deduced here from the SiC X grains.}
\label{fig:1}
\end{figure}

Figure \ref{fig:1} shows an extended plot of the variation of the $^7{\rm Li_\nu}/{^{11}{\rm B_\nu}}$ ratio as determined in \cite{Yoshida08} as a function of $\sin^2{(2 \theta_{1 3})}$.  The key point is that for $\sin^2{(2 \theta_{1 3})} > 10^{-3}$ there is a significant difference in this ratio between the normal and inverted hierarchies due to the effects of a higher $\nu_e$ energy in the normal hierarchy.  

The results of Refs.~\cite{Yoshida06a, Yoshida06b, Yoshida08}, however are only based upon a single 16.2 M$_\odot$ model and did not account for variation of $\nu$-process $^7$Li and $^{11}$B yields with progenitor mass  \cite{Heger05} and uncertainties in underlying thermonuclear reaction rates.  Indeed, it has been demonstrated  \cite{Austin11} that there are large uncertainties in the boron production due to the fact that uncertainties in the $3\alpha \rightarrow ^{12}$C and $^{12}C(\alpha,\gamma)^{16}$O nuclear reaction rates can lead to different core configurations and ultimately, differences in the  produced   $^{11}$B and $^{7}$Li by factors larger than the sensitivity to the neutrino mass hierarchy.  

  What is needed, therefore, is a proper statistical analysis of the uncertainties in all model parameters in order to determine whether $\nu$-process nucleosynthesis is a possible means to constrain the neutrino mass hierarchy.  The purpose of this paper is to explore such a possibility based upon a Bayesian analysis of the underlying uncertainties in the supernova models and measured meteoritic material.
  
\section{Meteoritic $\nu$-process Material}
All of this is well and good, but would be of no value without a measurement of $\nu$-process $^{7}$Li and $^{11}$B in supernova material enriched by the
$\nu$-process. The detection of such rare isotopes directly in supernova remnants would be exceedingly difficult.  Fortunately, however, there
is another way.  For many years it has been recognized that SiC  grains in carbonaceous chondrite meteorites can contain a component
of material freshly synthesized in a supernova and then trapped in the grain.  Most of the pre-solar grains appear to represent the ejecta from
AGB stars of 1-3 M$_\odot$; however a small fraction ($\sim 1\%$)  of them (the so-called X grains) can be attributed to  ejecta from core collapse supernovae \cite{Amari92, Nittler96, Hoppe00}.
The isotopic ratios of these  X grains exhibit $^{12}$C/$^{13}$C higher than solar  and $^{14}$N/$^{15}$N lower than normal.  They also have  enhanced $^{28}$Si.  These isotopic features are  all characteristic of core collapse supernovae.  Moreover, they contain decay products of radioactive 
isotopes synthesized in supernovae  such as $^{26}$Al ($t_{1/2} = 7 \times 10^5$ yr) and 
$^{44}$Ti ($t_{1/2} = 60$ yr), establishing that these grains indeed formed from encapsulated fresh supernova material \cite{Hoppe00}. 

In this context it is of particular note  that a recent study \cite{Fujiya11} of 1000 SiC grains from a 30 g sample of the Murchison CM2 chondrite \cite{Besmehn03} found 12 X grains that show resolvable anomalies in Li and/or B.
In particular, an average of the 7 best-case single grains indicated a  $2 \sigma$ detection of excess $^{11}$B relative to $^{10}$B (i.e. $^{11}$B/ $^{10}$B $= 4.68 \pm 0.31$ compared to the solar ratio of 4.03 \cite{Zhai96, Asplund09}) along with an enrichment of Li/Si and B/Si relative to solar.  As $^{11}$B is expected to be the main product of light element $\nu$-process nucleosynthesis \cite{Yoshida08}, this hints at  an average enrichment of $\sim 15\%$ of $\nu$-process $^{11}$B.  On the other hand they determined a slightly below solar average ratio of  $^{7}$Li/$^{6}$Li = $11.83 \pm 0.29$ compared to the meteoritic solar ratio of $12.06$ \cite{Seitz07}. However, the uncertainty is consistent with  some $^7$Li excess even at the $1 \sigma$ level.  Hence, this result can be used to set upper limits on the $^7$Li$_\nu/^{11}$B$_\nu$ production associated with these grains.

  Assuming that the observed grain  $^7$Li  and $^{11}$B abundances can be decomposed into  average solar plus $\nu$-process contributions, we then write:
  \begin{eqnarray}
  \frac{^7{\rm Li}}{^6{\rm Li}} &=& \frac{^7{\rm Li_\odot + ^7Li_\nu}}{^6{\rm Li_\odot}} \nonumber \\
  && = 12.06 + \frac{\rm ^7Li_\nu}{^6{\rm Li_\odot}} = 11.83 \pm 0.29 ~~,
  \end{eqnarray}
and similarly,
 \begin{eqnarray}
  \frac{^{11}{\rm B}}{^{10}{\rm B}} &=& 4.03 + \frac{\rm ^{11}B_\nu}{^{10}{\rm  B_\odot}} = 4.68 \pm 0.31 ~~,
  \end{eqnarray}
where we explicitly note that neither $^{6}$Li nor  $^{10}$B is
 significantly produced in the $\nu$-process \cite{Yoshida08}.
Next we adopt the solar ratios $^{7}$Li/$^{6}$Li $= 12.06 \pm 0.02$ \cite{Seitz07} and $^{11}$B/$^{10}$B$= 4.03 \pm 0.04 $ \cite{Zhai96}  consistent with carbonaceous chondrites.  We note that other $^{7}$Li/$^{6}$Li ratio values  exist in the literature based upon a combination of solar photospheric and meteoritic data, e.g. $^{7}$Li/$^{6}$Li$= 12.176$ \cite{Asplund09}, 12.33 \cite{Anders89}, 12.177 \cite{Lodders09}.  However, the value adopted here is the one  found in CI meteorites and hence the best representation of primitive solar system material.  

From Eqs.~(1) and (2) we can deduce 
\begin{equation}
  \frac{^7{\rm Li_\nu}}{^{11}{\rm B_\nu}} \biggl( \frac {^{10}{\rm B}_\odot}{^{6}{\rm Li}_\odot} \biggr)_{X} = -0.35  \pm 0.48
 ~~ ~~,
  \end{equation}
where the subscript $X$ denotes the $^6$Li/$^{10}$B ratio appropriate to the X-grain samples under study.  Although it is probably safe to adopt the solar isotopic ratios for the grain contaminant, elemental ratios are not appropriate as there is no guarantee that the non-$\nu$-process Li/B in the X grains is identical to the solar value due to the different volatilities of Li and B. Fortunately, however this ratio can be determined directly for the grains from the measured elemental ratios in \cite{Fujiya11}.  The average for the 7 single grains is Li/Si$= 8.27 \pm 0.06 \times 10^{-5}$  and  B/Si$= 4.13 \pm 0.11 \times 10^{-5}$.

From this we write:
\begin{equation}
\biggl[\frac{  (^7{\rm Li}_\odot+ ^7{\rm Li}_\nu)/^6{\rm Li}_\odot + 1}{(^{11}{\rm B}_\odot + ^{11}{\rm B}_\nu)/^{10}{\rm B}_\odot + 1}\biggr]_X = 2.00 \pm 0.04 \biggl(\frac{^{10}{\rm B_\odot}}{^6{\rm Li_\odot}} \biggr)_{\rm X}  ~~.
\end{equation}

Then inserting the X grain measured isotopic ratios given in  Eqs. (1) and (2), we deduce an average X-grain ratio of  ($^{6}$Li$_\odot/^{10}$B$_\odot)_X = 0.89 \pm 0.06$.  Inserting this into Eq.~(3) leads to $^7$Li$_\nu/^{11}$B$_\nu = -0.31 \pm 0.42$ and  the following upper limits on the X-grain $\nu$-process isotopic ratio:
\begin{eqnarray}
\frac {^7{\rm Li_\nu}}{^{11}{\rm B_\nu}} & < & 0.53 (2 \sigma ~ 95\%~{\rm C.L.})  \nonumber \\
 & < & 0.95 (3 \sigma~99.7\%~{\rm C.L.})~~.
 \end{eqnarray}

The width of the shaded region on Fig.~\ref{fig:1} shows the $2 \sigma$ (95\% C.L.)  determination of  $\sin^2{(2 \theta_{1 3})}$  from the Daya Bay analysis.  The top of the shaded regions is the $2 \sigma$  upper limit to the $^7$Li$_\nu$/$^{11}$B$_\nu$ ratio from the meteoritic SiC X grain analysis.   

 \section{Bayesian Analysis}

Even though the meteoritic data only provides an  upper limit, one can use Bayesian statistics to quantify the degree to which the  information in this evidence supports an inverted or normal hierarchy neutrino mass hierarchy.
In particular, one can use Bayes'  theorem \cite{HowsonUrbach,Gregory} to ascertain the probability $P(M_i \vert D)$ that a given model $M_i$ ($i=$inverted or normal mass hierarchy) is true based upon a data set $D$ (the meteoritic evidence).  

According to Bayes theorem we have:
\begin{equation}
  P(M_i\vert D )=\frac{ P(D \vert M_i) P(M_i) }{\sum_j P(D \vert M_j) P(M_j) }~~,
\label{bayes}
  \end{equation} 
where $P(D\vert M_i) $ is a likelihood function giving the probability of the data to be reproduced by model $M_i$ (including all uncertainties in the model parameters) and  $P(M_i) $ is the prior probability (e.g. for our purposes we take to be $50/50$) for an inverted versus normal hierarchy.  The  sum in the denominator is over all possible models, which in our case is just the inverted plus normal neutrino mass hierarchy.  We will ignore here the possibility of no oscillations as $\sin^2{2 \theta_{1 3}} =0$ is now ruled out at the level of 5.2$\sigma$ \cite{DayaBay}.

The key to this analysis is to identify the likelihood functions  $P(D\vert M_i)$.  To do this we separate the likelihood into the probability distribution $P(a_k\vert M_i)$ of the parameters $a_k$ of the model  and the probability $P(E,Z,D \vert M_i, a_ {k})$ that a given model $i$ with a set of parameters $a_ {k}$ produces a result $Z$ that agrees with the observed data $D$ that has an experimental error $E$. Then we have 
  \begin{eqnarray}
  P(D \vert M_i) = \int dE dZ   da_k P(E,Z,D \vert M_i, a_k) P(a_k \vert M_i) ~, \nonumber \\
 = \int dE dZ   da_k P(D \vert M_i, a_k,E,Z) P(Z,E  \vert M_i, a_k)P(a \vert M_i)~.  
 \label{bayesint}
  \end{eqnarray}
  
  Now for each set of parameters $a_ {k}$ there is a unique prediction $Z_i$ for model $M_i$, so $P(Z\vert M_i, a_k) =  \delta[Z-Z_i(M_i,a_k)]$.  
  Moreover, the probability that  a given $Z_i$   agrees with the observed data plus error  $D +E$  is given by the error distribution  in the data (taken here to be gaussian).
  For the present application $D \pm \sigma_E = -.31 \pm 0.42$ as deduced from the meteoritic data,  so that we have
  \begin{equation}
  P(E \vert Z_i ( M_i, a_k) ) = \frac{1}{\sqrt{2 \pi} \sigma_E} \exp[ -(Z_i - D)^2/2 \sigma_E^2)~~.
  \end{equation}

  Next we need to specify the probability distributions $P(a_k \vert M_i)$ for the parameters $a_k$ of the models.
  For our purposes we identify 5 parameters, none of whose prior probability distributions depend upon whether the hierarchy is normal or inverted.  So Eq.~(\ref{bayesint}) becomes a five-dimensional integral over the probability distributions  of each parameter weighted by the gaussian experimental error.  These parameters are: $\sin^2{2 \theta_{1 3}}, T_\nu, R_{3 \alpha}, R_{12C\alpha},M_{prog}$ as listed in Table \ref{tab:1}.  
  
  The parameter likelihood functions were decided in the following way. 
 We adopt the Daya Bay result \cite{DayaBay} of  $\sin^2{2 \theta_{1 3}} = 0.092$  as it has the smallest systematic error and is close to the weighted average of the three best recent determinations.  Since the Daya Bay result is dominated by  statistical error, it is sufficient for our purpose to simply add the statistical and systematic  errors in quadrature.  We then adopt an overall gaussian probability distribution for $\sin^2{2 \theta_{1 3}}$ with a standard deviation of $\sigma = 0.017$.
 
   For the rate of the $3 \alpha \rightarrow ^{12}$C reaction, as in \cite{Austin11} we consider an overall  multiplier $R_{3\alpha}$, where a value of unity corresponds to the standard evaluation \cite{CF88} with a 12\% gaussian error.  Similarly, we consider a multiplier $R_{C12\alpha}$ for the $^{12}$C$(\alpha, \gamma)^{16}$O reaction, with a best value \cite{Buchmann} of $R_{C12\alpha} = 1.2 \pm 0.25$ times the evaluation in Ref.~\cite{CF88}.  
   
   The probability distribution for a  progenitor star to have a  mass $m$ is given by the initial mass function \cite{Scalo} $\phi(m) = m^{-x}$, with x = 2.65, and the normalization cancels out in this and all parameter probability distributions.  
   The range of progenitor masses that can undergo core-collapse supernovae, however is limited to a range of 10 to 25 M$_\odot$ \cite{Heger05}.  Below 10 M$_\odot$ the star does not form an iron core.  Above 25 M$_\odot$ a black hole probably forms rather than a neutron star.  This interrupts the flux of neutrinos from the core so that there is no $\nu$-process.
  
  For the neutrino temperature, the situation is a little more complicated.  There are in fact six temperatures corresponding to the electron, muon, and tau neutrinos plus their anti-particles.  Although the cross sections are roughly proportional to the neutrino temperature squared, this scaling cancels to first order when taking  the ratio of yields. Even so, there is a small residual sensitivity of the final $^{7}$Li$/^{11}$B ratio to the neutrino temperatures which must be included.
    
    Fortunately, one  can constrain  phenomenologically the relative neutrino temperatures and their possible variations.  For example, a recent detailed analysis \cite{Hayakawa10} of the relative
production of $^{180}$Ta and $^{138}$La in the $\nu$-process fixes the
temperature of the electron neutrinos and their anti-particles to be around 4 MeV.
On the other hand, successful r-process nucleosynthesis conditions require a neutron-rich environment
in the SN neutrino-driven winds.  This requires  that the electron neutrino temperature be
slightly lower than that of the anti-electron neutrino temperature \cite{Yoshida04},
i.e. $T_{\nu_e} = 3.2$ MeV for $T_{\bar \nu_e} \approx 4.0$ MeV.
Additionally, the observed galactic chemical evolution of boron and the meteoritic $^{11}$B/$^{10}$B ratio
have been  shown \cite{Yoshida05} to constrain the $\nu_\mu$ and $\nu_\tau$ neutrino temperature 
 to be about  5 MeV with an allowed range of $T_{\nu_{\mu,\tau}} = 4.3$ to $6.5$ MeV.
This corresponds to Model ST in Ref.~\cite{Yoshida08}.  We therefore adopt these as the best set of average neutrino temperatures.   We then consider the range of models (1,2,LT and ST) studied in \cite{Yoshida08} as a suitable representation of the range of variation of the $^{7}$Li$/^{11}$B ratio with neutrino temperature in  core collapse models.  These models incorporate the following ranges: $T_{\nu_{e}} = 3.2-4.0$ MeV;  $T_{\bar \nu_e} = 4.0-5.0$ MeV; $T_{\nu_{\mu,\tau}} = 5.0-6.5$ MeV.  The span of $^{7}$Li$/^{11}$B ratios from this range is represented by the two sets of solid lines for each scenario in Figure \ref{fig:1}.  We treat this as a top-hat distribution in the integration of Eq.~(\ref{bayesint}).  
The sensitivity of the  neutrino process to these models and also to variations in the neutrino cross sections  has been carefully studied in \cite{Yoshida08} and does not significantly impact the discussion here.

Another sensitivity arises from the swapping  of neutrino energy
spectra due to the self-interaction among neutrinos \cite{Pastor02, Balantekin05, Fogli07, Duan08}. 
It is well known that no one  has yet  succeeded  in solving exactly the many-body problem of three flavor oscillation due to the neutrino self-interaction.  Nevertheless, as is demonstrated by Fogli et al. \cite{Fogli07}, several cases of complete swapping of the high energy neutrinos between $e$- and $\mu$- plus $\tau$-neutrinos (or between their anti-particles)  should mimic the flavor oscillation induced by an adiabatic MSW high-density resonance:  See Figs. 5 and 8 of Ref. \cite{Fogli07}. This illustrates the maximum effect on the charged current  interactions of the neutrino-process nucleosynthesis of interest to the present study.  Hence, even should
that effect  be included in our adopted neutrino temperatures, the present
discussion would not change drastically.  This is  because the swapping effect is
expected to result in  the same  isotopic ratios, i.e.  $^7$Li$/^{11}$B $\sim 0.8$ for
a normal hierarchy and $\sim 0.6$ for an inverted hierarchy even for a small mixing
angle $\theta_{13} < 0.001$ which is excluded by current experiments.

\vspace{.2in}
\begin{table}
\centering
\caption{Parameter likelihood functions $P(a_{k} \vert M_i)$.}
\begin{tabular}{| c | c | c | c | c |c|} \hline
 Parameter $a_k$& prior  & & &      reference  \\ \hline
$\sin^2{2 \theta_{1 3}}$ &  $e^{-(x-x_0)/2 \sigma_x^2}$  & $x_0 = 0.92$ & $\sigma_x  = 0.017$ &   \cite{DayaBay} \\ \hline 
$R_{3\alpha}$ & $e^{-(x-x_0)/2 \sigma_x^2}$  & $x_0 = 1.0$ & $\sigma_x  = 0.12$ &   \cite{CF88} \\ \hline 
$R_{12C\alpha}$ & $e^{-(x-x_0)/2 \sigma_x^2}$  & $x_0 = 1.2$ & $\sigma_x  = 0.25$ &   \cite{Buchmann} \\ \hline 
$M_{prog}($M$_\odot$) & $m^{-2.65} $ & $m_{min} = 10$ & $ m_{max} = 25$ & \cite{Scalo} \\ \hline
$T_\nu$(MeV) & Top hat & $T_\nu = 3.2-6.5$ & (see text)  &  \cite{Yoshida08} \\ \hline  
\end{tabular}
\label{tab:1}
\end{table} 

\vspace{-.1in}

So now the likelihood for an inverted vs. a normal hierarchy becomes a five dimensional integral over the parameters listed in Table \ref{tab:1}.
 \begin{equation}
 P(M_i \vert D) = \frac{\int d^5a_k e^{-(D-Z_i)^2/2 \sigma_E^2} \Pi_k P(a_k\vert M_i)}
 {\sum_i \int d^5a_k  e^{-(D-Z_i)^2/2 \sigma_E^2} \Pi_k P(a_k\vert M_i)} ~~.
 \label{5dint}
 \end{equation}

The final task is then to specify the model prediction $Z_i(a_k,M_i)$.  In this we are aided by the fact that the 16.2 M$_\odot$ model calculation of \cite{Yoshida08} agrees well with the 15 M$_\odot$ models of \cite{Heger05,Woosley95} and also those of \cite{Austin11}.  Moreover, the relative effect of the oscillations on the $^7$Li/$^{11}$B ratio depends only on the mixing angle and not on the detailed abundance ratio in the absence of oscillations. Hence, we can utilize the results of \cite{Yoshida08} for the dependence of the $^7$Li/$^{11}$B ratio as a function of $\sin^2{2 \theta_{1 3}}$, as shown in Figure \ref{fig:1}.  The temperature dependence is given as a linear offset of the model calculations for each value of $\sin^2{2 \theta_{1 3}}$.  All other model parameter dependences can be taken as a correction factor to the  15 M$_\odot$  calculation.  That is we write:
 \begin{eqnarray}
   Z_i(\theta_{1 3}, T_\nu, R_{3 \alpha} T_{12C\alpha}, M) &=& [Z(\theta_{1 3}) + \Delta Z(T_\nu)]  \\
   &\times& f_{R3\alpha} \times f_{R12C\alpha} \times f_M~, \nonumber
   \end{eqnarray}
   where $Z(\theta_{1 3}) + \Delta Z(T_\nu)$ is the 16.2 M$_\odot $ model calculation of \cite{Yoshida08}.  The temperature correction $\Delta Z(T_\nu)$ is the top-hat distribution of results shown on Figure \ref{fig:1} based upon models with different neutrino temperatures \cite{Yoshida08}.

\begin{figure}[htb]
\includegraphics[width=3.5in,clip]{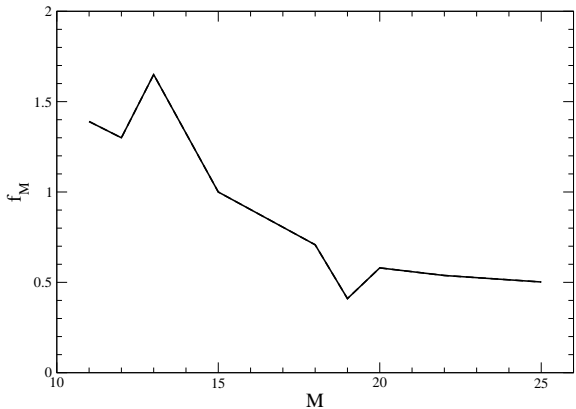} 
\caption{ Correction factor for the $\nu$-process $^7$Li/$^{11}$B ratio relative to the 15 M$_\odot$ model as a function of  progenitor mass (deduced from \cite{Woosley95}).}
\label{fig:2}
\end{figure}

\begin{figure}[htb]
\includegraphics[width=3.5in,clip]{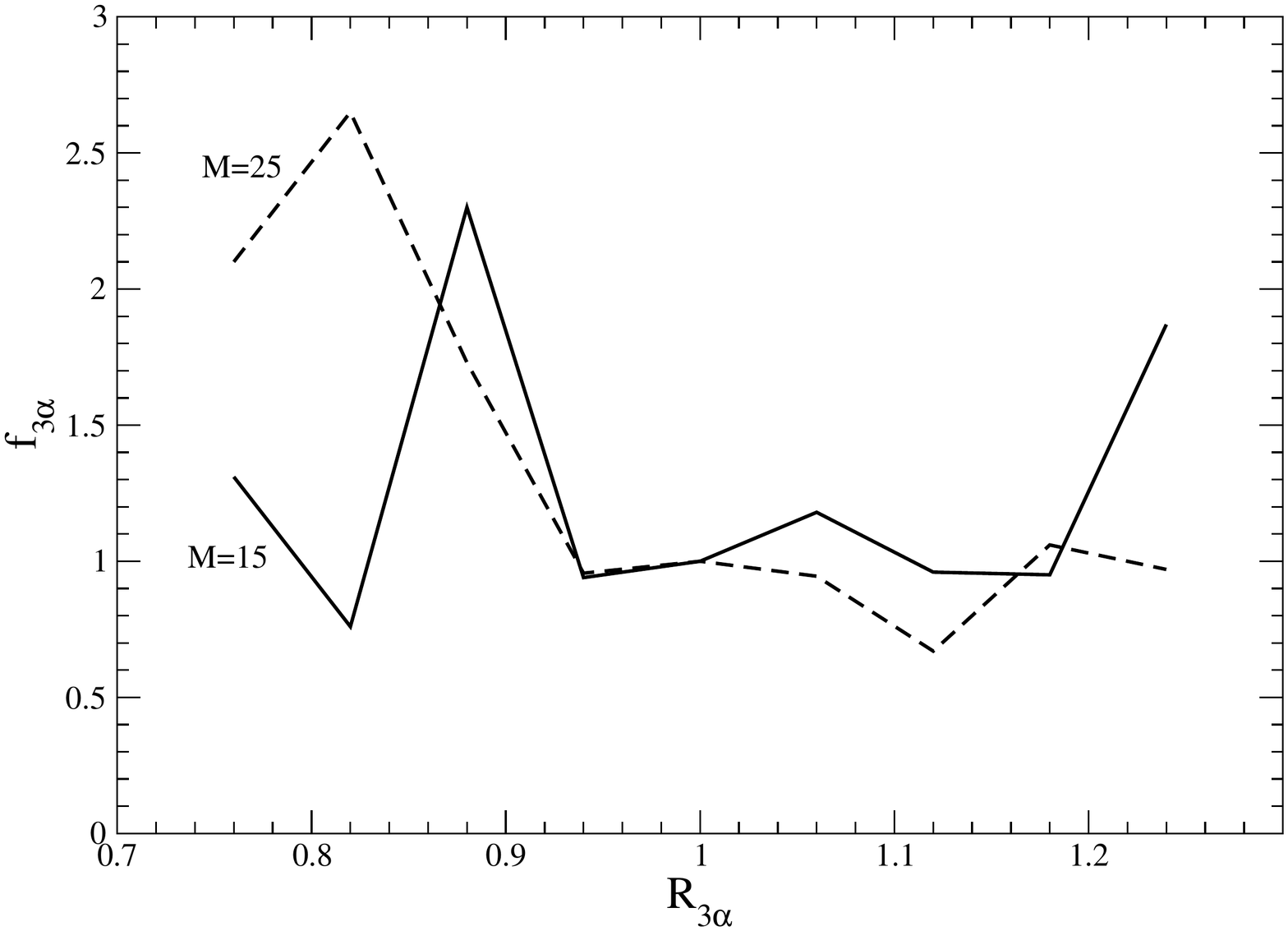} 
\caption{ Correction factor for the  $^7$Li/$^{11}$B ratio relative to the adopted rate ($R_{3 \alpha} =1$) as a function of  the $3 \alpha \rightarrow ^{12}$C  nuclear reaction rate (deduced from \cite{Austin11}).}
\label{fig:3}
\end{figure}

\begin{figure}[htb]
\includegraphics[width=3.5in,clip]{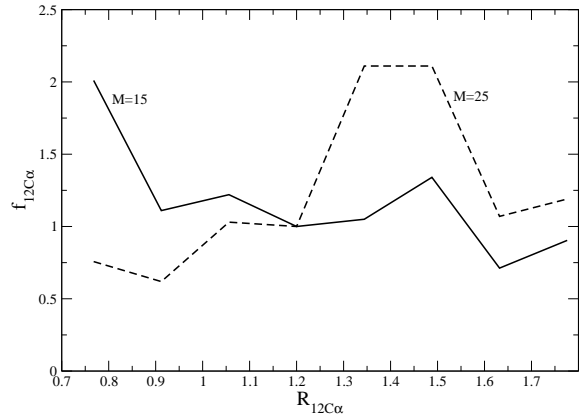} 
\caption{ Correction factor for the  $^7$Li/$^{11}$B ratio relative to the adopted rate ($R_{12C\alpha} =1.2$) as a function of  the $^{12}$C$(\alpha,\gamma)^{16}$O nuclear reaction rate (deduced from \cite{Austin11}).}
\label{fig:4}
\end{figure}

The relative ratios as a function of the mixing angle should not change much as the progenitor mass changes because this will be fixed by the same resonance condition independently of the other parts of the star.  Hence, correction factors for the dependence of the $^7$Li/$^{11}$B ratio as a function of progenitor mass can be obtained from the $\nu$-process (without oscillations) calculations of \cite{Woosley95}.  These factors are plotted in Figure \ref{fig:2}.  Similarly, correction factors as a function of the $R_{3 \alpha}$ and the $R_{12C \alpha}$ as a function of progenitor mass can be extracted  from \cite{Austin11}.  These are shown in Figures \ref{fig:3} and \ref{fig:4}.  Unfortunately results in Ref.~\cite{Austin11} are only given for progenitor masses of 15 and 25 M$_\odot$.  For our purposes we have assumed constant production factors outside that mass range and linearly interpolated production factors inside that mass range.
 
 Based upon these assumptions we have preformed a numerical evaluation of the five dimensional integral given in Eq.~(\ref{5dint}).  When completed we found that a prior equal probability (50/50) for an inverted vs. a normal mass hierarchy becomes a 74/26 probability based upon the adopted parameters, their prior probabilities, and influence on the model results.  Hence, although presently there are large uncertainties in this approach, based upon the present analysis there appears  to be a slight preference for an inverted neutrino mass hierarchy.

\section{Conclusion}

 In summary, we have shown that an analysis of SiC X grains enriched in $\nu$-process material has the potential to solve the neutrino mass hierarchy problem in spite of the large model uncertainties and the difficulties in determining the abundance of $^7$Li$_\nu$ and $^{11}$B$_\nu$.  The  fact that $^7$Li is produced more efficiently than $^{11}$B in the case of a normal hierarchy due to the higher $\nu_e$ energy after oscillations for finite  $\theta_{1 3}$  hints that a combination of  accelerator limits, SiC grain analysis, and an analysis of supernova models  might point toward a solution of  the neutrino hierarchy problem.  

There are of course, a large number of caveats here.  For one, we have used a rather crude approximation of the model dependence of the $^7$Li/$^{11}$B ratio on various parameters from published results in the literature.  Clearly a more exhaustive self consistent set of model calculations would be desirable.  Nevertheless, the simple analysis applied here  at least demonstrates that it might be worthwhile to under take such an extensive study.

Another caveat is that of course there are many uncertainties in the meteoritic analysis.  For example the present  analysis assumes that the $^7$Li$_\nu$ and $^{11}$B$_\nu$ condensed into the SiC grain in a way that preserves  the $^7$Li$/^{11}$B
ratio in the SiC condensation site of the SN ejecta.  However, because of  elemental fractionation
between the SN ejecta and the SiC grain, the $^7$Li/$^{11}$B ratio in  the X grain might be
different from the true value of the ejecta. 

We also note that the analysis here is based upon an average of only  7 single X  grains and provides only an upper limit on  the   $^7$Li$_\nu/^{11}$B$_\nu$ ratio because there is no detection of $\nu$-process $^{7}$Li (although some individual grains do show enhanced $^7$Li/$^6$Li \cite{Fujiya11}).  Both mass hierarchies are thus constrained.  Nevertheless, we have shown that without prior preference for either hierarchy,  the inverted hierarchy  is  preferred over the normal hierarchy by a factor of $\sim 3/1$.   Clearly, more analysis of $\nu$-process material in SiC meteoritic grains is warranted.  An analysis of even a few more grains could provide a detection of $\nu$-process $^7$Li and substantially improve these limits.  Such a  detection  would provide both upper and lower limits to the $^7$Li$_\nu/^{11}$B$_\nu$  ratio and confirm this means to determine the neutrino mass hierarchy.

\begin{acknowledgments}
Work at the University of Notre Dame is supported (GJM)
by the U.S. Department of Energy under
Nuclear Theory Grant DE-FG02-95-ER40934  and (JPB) by the UND Center for Research Computing.
Work at NAOJ is supported in part by the Grants-in-Aid for
Scientific Research of the JSPS (20244035), Scientific Research on Innovative Area of MEXT (2010544), and the Heiwa Nakajima Foundation.
\end{acknowledgments}


\begin{references}

\bibitem{SuperK05} Y.~Ashie, et al.~(Super-Kamiokande Collaboration), Phys. Rev.{\bf D71}, 112005 (2005).
%
\bibitem{SuperK06} J.~Hosaka, et al.~(Super-Kamiokande Collaboration), Phys.Rev.~{\bf D73}, 112001 (2006).
%
\bibitem{SNO} B.~Aharmim, et al.~(SNO Collaboration), Phys. Rev.~{\bf C72}, 055502 (2005),.
%
\bibitem{K2K} M.~H.~Ahn, et al.~(K2K Collaboration), Phys. Rev. {\bf D74}, 072003 (2006)2.
%
\bibitem{KamLAND} S.~Abe, et al.~(KamLAND Collaboration), Phys.~Rev.~Lett. {\bf 100}, 221803 (2008),.
%
\bibitem{MINOS}P.~Adamson, et al.~(MINOS Collaboration),  arXiv:1108.0015 [hep-ex].
%
\bibitem{DayaBay} F.~P.~An,~et al. (Daya Bay Collaboration) archive:1203.1669 [hep-ex].
%
\bibitem{RENO} J.~K.~Ahn, et al. (RENO Collaboration),  hep-ex arXiv:1204.0626 (2012).
%
\bibitem{Chooz} Y.~Abe, et all (Double Chooz Collaboration), Phys. Rev. Lett. 108, 131801 (2012). 
%
\bibitem{T2K} K.~Abe, et al.~(T2K Collaboration), Phys. Rev. Lett.~{\bf 107}, 041801  (2011).
%
\bibitem{Yoshida04} T.~Yoshida, et al., Astrophys. J. {\bf 600}, 204 (2004).
%
\bibitem{Yoshida05}  T.~Yoshida, T.~Kajino, and D.~H.~Hartmann, Phys. Rev. Lett. {\bf 94}, 231101 (2005).
%
\bibitem{Yoshida06a}T.~Yoshida, T.~Kajino, H.~Yokomakura, K.~Kimura, A.~Takamura,  and
D.~H.~Hartmann, Phys. Rev. Lett., 96, 091101 (2006).
%
\bibitem{Yoshida06b}
T. Yoshida, T.~Kajino, H.~Yokomakura, K.~Kimura, A~Takamura and 
D.~H.~Hartmann,   Astrophys. J., {\bf 649}, 319 (2006).
%
\bibitem{Yoshida08}
T.~Yoshida, et al., Astrophys. J., 686, 448 (2008).
%
\bibitem{Fujiya11} W.~ Fujiya, P.~Hoppe, and U.~Ott, Astrophys. J. Lett., {\bf730}, L7 (2011).
%
\bibitem{Austin11}S.~M. Austin, A.~Heger and C.~Tur, Phys. Rev. Lett., {\bf 106}, 152501 (2011).
%
\bibitem{Heger05} A.~Heger, E.~Kolbe, W.~C.~Haxton, K.~Langanke, G.~Mart{\'i}nez-Pinedo  and S.E.~Woosley, Phys. Lett. {\bf B606}, 258 (2005).
%
\bibitem{Woosley95} S.~E.~Woosley and T.~A.~Weaver, Astrophys. J. Suppl.,  {\bf 101}, 181 (1995).
%
\bibitem{HowsonUrbach} C.~Howson \& P.~Urbach, \emph{Scientific Reasoning: The Bayesian Approach}, 2nd ed., Open Court, Chicago (1993).  
%
\bibitem{Gregory} P.~C.~Gregory, \emph{Bayesian 
Logical Data Analysis for the Physical 
Sciences:  A Comparative Approach with 
Mathematica (R) Support}, (Cambridge U. Press, Cambridge, 2005).  
%
\bibitem{Domogatsky78}
G. V.~Domogatsky, R. A.~Eramzhyan, and D.~K.~Nadyozhin, 
in {\it Proc. Int. Conf. on Neutrino Physics and Neutrino Astrophysics},
ed. M. A. Markov, et al.,
(Moscow: Nauka),  115, (1978).

\bibitem{Woosley90}S. E. Woosley et al., Astrophys. J. {\bf 356}, 272 (1990).
%
\bibitem{Dighe00} A. S. Dighe, and A.Y. Smirnov, Phys. Rev. {\bf D62}, 033007
(2000); K. Takahashi, M. Watanabe, K. Sato, and T. Totani, Phys. Rev. {\bf D64}, 093004
(2001); K. Takahashi, and K. Sato, Prog. Theor. Phys.
{\bf 109}, 919 (2003); K. Takahashi, et al., Astropart. Phys.
{\bf 20}, 189 (2003).
%
\bibitem{Hayakawa10} T.~Hayakawa, T.~Kajino, S.~Chiba and   G.~J.~Mathews,  Phys. Rev., {\bf  C81}, 052801 (2010); T.~Hayakawa, P.~Mohr, T.~Kajino, S.~Chiba and G.~J.~Mathews, Phys. Rev., {\bf  C82}, 058801 (2010).
%
\bibitem{Amari92} S.~Amari, P.~Hoppe, E.~Zinner and R.~S.~Lewis, Astrophys. J. Lett., {\bf 394}, L43 (1992). 
%
\bibitem{Nittler96} L.~R.~Nittler, S.~Amari, E.~Zinner, S.~E.~Woosley and R.~S.~Lewis,  Astrophys. J. Lett.,
{\bf 462}, L31 (1996).
%
\bibitem{Hoppe00} P.~Hoppe, et al., Meteoritics and Planetary Sci.,  {\bf35}, 1157 (2000).
%
\bibitem{Besmehn03} A.~Besmehn, and P.~Hoppe, Geochim.~Cosmochim.~Acta, {\bf 67}, 4693 (2003).
%
\bibitem{Zhai96} M.~Zhai, et al., Geochim. Cosmochim. Acta, ,  {\bf60}, 4877 (1996).
%
\bibitem{Asplund09} M.~Asplund, N.~Grevesse, A.~J.~Sauval, and P.~Scott, Ann.~Rev.~Astron.~Astrophys., {\bf 47}, 481 (2009).
%
\bibitem{Seitz07} H.-M.~Seitz, et al., Earth and Planet. Sci. Lett., {\bf 260}, 582 (2007).
%
\bibitem{Anders89} E.~Anders, and N.~Grevesse, Geochim. Cosmochim. Acta {\bf 53}, 197 (1989).
%
\bibitem{Lodders09} K. Lodders, H.~Palme \& H.~P.~Gail,   In {\it Landolt-Bšrnstein, New Series,} Vol. VI/4B, J.E. TrŸmper (ed.), 
Springer-Verlag, pp. 560-630, (2009).
%
\bibitem{CF88} G. R. Caughlan and W. A. Fowler, At. Data Nucl. Data Tables {\bf 40}, 283 (1988).
%
\bibitem{Buchmann} L. R. Buchmann, Astrophys. J. {\bf 468}, L127 (1996); {\bf 479}, L153(E) (1997).
%
\bibitem{Scalo} J. M. Scalo, Fundam. Cosm. Phys. {\bf 11}, 1 (1986).
%
\bibitem{Pastor02}S.~Pastor and G.~Raffelt, Phys. Rev.~Lett., {\bf 89}, 191101 (2002).
%
\bibitem{Balantekin05} A.~B.~Balantekin and H.~Y\" uksel, New. J.~Phys., {\bf 7}, 51 (2005).
%
\bibitem{Fogli07} G.~L.~Fogli, et al., J.~Cosmol.~Astropart.~Phys., {\bf 12}, 010, 2007).
%
\bibitem{Duan08} H.~Duan, G.M.~Fuller, J.~Carlson, and Y.Z.~Qian, Phys.~Rev.~Lett., {\bf 100}, 021101 (2008).

\end{references}
\end{document}